\documentclass [a4paper]{JHEP3}
\usepackage{amssymb,graphicx}
\usepackage{amsmath}
\usepackage{epsfig}
\title{%
\protect\vspace{5mm}
On brane-induced gravity in warped backgrounds
}

\def\e{{\rm e}}

\def\d{\partial}
\def\l{\left(}
\def\r{\right)}

\newcommand{\be}{\begin{equation}}
\newcommand{\ee}{\end{equation}}
\newcommand{\bea}{\begin{eqnarray}}
\newcommand{\eea}{\end{eqnarray}}
\newcommand{\bg}{\begin{gather}}
\newcommand{\eg}{\end{gather}}
\newcommand{\bseq}{\begin{subequations}}
\newcommand{\eseq}{\end{subequations}}

\author{S.L.~Dubovsky$^a$ and M.V.~Libanov$^{a,b}$\\
 $^a$
Institute for Nuclear Research of
         the Russian Academy of Sciences,\\  60th October Anniversary
  Prospect, 7a, 117312 Moscow, Russia  \\
$^b$ Service de Physique Th\'{e}orique, CP 225,\\
  Universit\'{e} Libre de Bruxelles, B--1050, Brussels, Belgium\\
E-mail: \email{sergd@ms2.inr.ac.ru, ml@ms2.inr.ac.ru} }
\preprint{ULB-TH-03-30\\
INR/TH-2003-7}
\abstract{
We study whether modification of gravity at large distances is possible in
warped backgrounds with two branes and a brane-induced term localized on
one of the  branes. We find that there are three large regions in the
parameter space where the theory is weakly coupled up to high energies. In
one of these regions gravity on the brane is four-dimensional at
arbitrarily large distances, and the induced Einstein term results merely
in the renormalization of the 4d Planck mass. In the other two regions the
behavior of gravity changes at ultra-large distances; however, radion
becomes a ghost. In parts of these regions, both branes have positive
tensions, so the only reason for the appearance of the ghost field is the
brane-induced term. In between these three regions, there are domains in
the parameter space where gravity is strongly coupled at
phenomenologically unacceptable low energy scale.
}
\begin{document}
\section{Introduction} Emerging evidence for the accelerated expansion of
the Universe triggered interest in the non-standard theories of gravity in
which gravitational interactions get modified at length scales larger than
a certain critical value $r_c$, which is assumed to be of the order of the
present Hubble length. The motivation is that the unconventional behavior
of gravity at cosmologically large length scales may either screen the
effect of the huge vacuum energy or even be at the origin of cosmic
acceleration if the vacuum energy is exactly zero.

Naively, the most straightforward way to modify gravity at the distance
scale $r_c$ is to give graviton a mass $m_g\sim r_c^{-1}$. However, it is
known for a long time \cite{vanDam:1970vg} that the massive gravity does
not reproduce the conventional Einstein gravity at the linearized level in
the limit $m_g\to 0$ due to the different tensor structures of the
propagators for  the massive and massless  gravitons. It was argued
\cite{Vainshtein:sx} that this problem may be cured by taking into account
non-linear effects. However, recent study \cite{Arkani-Hamed:2002sp}
demonstrated that any four-dimensional theory of a single massive graviton
either exhibits ghost fields or loses its predictive power in the UV due
to strong quantum effects at the scale $\Lambda=\l M_{Pl}/r_c^2\r ^{1/3}
$. If $r_c$ is of the order of the Hubble distance $H^{-1}\sim 10^{28}$~cm
then $\Lambda \sim 10^{-8}$~cm$^{-1}$ which is unacceptably low from the
phenomenological point of view. This strong interaction is due to the fact
that when the graviton mass term is added, extra scalar and vector degrees
of freedom become dynamical and strongly interacting, and as a result
there is no smooth limit $m_g\to 0$ (a notable exception  from this rule
is massive gravity in the adS space \cite{Porrati:2000cp}).

This situation is somewhat reminiscent of what happens with non-Abelian
Yang--Mills theory after gauge invariance is explicitly broken by the mass
term, so one is tempted to try to find an analogue of the Higgs mechanism
providing a predictive UV completion of massive gravity.

Models with infinite extra dimensions (``brane worlds'') appear to be a
natural framework for modifying gravity at large distance scales. Indeed,
in  brane world models four-dimensional physics is reproduced due to the
presence of KK modes with wave-functions localized near our brane. A
number of models \cite{Charmousis:1999rg}-\cite{Dvali:2001ae} have been
 proposed where four-dimensional graviton is actually a quasilocalized
state with a tiny but finite width $\Gamma$ (and, possibly, mass). In
these models one may expect that gravity is four-dimensional at distances
shorter than $r_c\sim \mbox{min}\{m^{-1},\Gamma^{-1}\}$ and
multi-dimensional at longer scales. Also one may hope that the presence of
a continuum of light bulk modes gives rise to non-locality, which is
strong enough to get around the no-go result obtained in the
four-dimensional theory.

Unfortunately, this is not the case as yet. In the five-dimensional model
of Refs. \cite{Charmousis:1999rg,Gregory:2000jc}, where   graviton is
quasilocalized due to the warped geometry of the bulk space, a ghost field
was found \cite{Pilo:2000et}. Its presence leads to the correct tensor
structure of the graviton propagator at intermediate scales but makes the
whole theory inconsistent. An interpretation is that this ghost is due to
the presence of the dynamical negative tension brane in the model that
violates the weakest energy condition.

In another class of models \cite{Dvali:2000hr,Dvali:2000xg,Dvali:2001ae}
 the bulk space is flat and the graviton is quasilocalized due to the
presence of the four-dimensional Einstein term on the brane. It has been
argued that this term may be induced by the loops of particles localized
on the brane. Naively, one does not expect ghosts in this setup as the
positive energy condition is not violated. Surprisingly, tachyonic ghost
field was found \cite{Dubovsky:2002jm} in the model with the number of
extra dimensions $N$ greater than one. In fact, $N>1$ models are somewhat
special, because they exhibit singularity in the thin brane limit and need
 some regularization to resolve this singularity. In Ref.
\cite{Dubovsky:2002jm} a particular regularization suggested in Ref.
\cite{Dvali:2001ae} was used. One may consider alternative regularizations
(see, e.g. Refs. \cite{Dubovsky:2001pe,Kolanovic:2003am}), where ghost
fields are absent; however, then one again faces strong coupling at the
unacceptably low energy scale.

The existence of ghosts in a setup where the positive energy condition is
maintained may appear somewhat puzzling.  To understand the situation, it
is instructive to study the case of one extra dimension which is free of
singularities. The flat space model \cite{Dvali:2000hr} with induced
Einstein term on the brane was studied in much detail in Refs.
\cite{Luty:2003vm,Rubakov:2003zb}. This model does not possess ghosts;
however there is  strong coupling like in the four-dimensional theory with
massive graviton. The strongly interacting mode is basically a brane
bending mode, so one may suspect that strong coupling is related to the
fact that the brane has zero tension.  Thus one is naturally led to
consider models with non-zero brane tension and, in static situation,
warped bulk space.  The study of models with warped bulk space and induced
term on the brane was started in Ref. \cite{Luty:2003vm} (for earlier
work, where the case of a scalar field was considered, see, e.g., Ref.
\cite{Kiritsis:2002ca}). There the brane was assumed to be at the fixed
point of the $Z_2$ orbifold. In particular, it was shown that in the
second Randall--Sundrum model with a single positive tension brane, the
induced term does not change physics in the regime of validity of the
four-dimensional effective theory and merely renormalizes the value of the
Plank mass on the brane. However, gravity is not modified at large
distances in that model.

In this paper we continue the study of models with warped bulk and induced
Einstein term on the brane. We drop the assumption of the $Z_2$ symmetry
across our brane and consider the effect of the Einstein term on one of
 the branes in the models of the Lykken--Randall type
\cite{Lykken:1999nb}.  We demonstrate that the behavior of these models is
in some respect similar to the behavior of massive four-dimensional
gravity. Namely, there are regions in the parameter space where the theory
is weakly coupled up to high energies. In one of these regions the induced
term is localized on the Planck brane. In this region the induced Einstein
term results merely in the renormalization of the 4d Planck mass (and the
absence of 5d behavior in the UV), so that gravity is not modified at
large distances. In the other regions the behavior of gravity changes at
ultra-large distances; however, radion becomes a ghost. Different regions
in the parameter space are separated by the domains where gravity is
strongly coupled at unacceptably low energy scale. It is worth noting that
there is a region in the parameter space where all branes have positive
tensions, but the ghost field is still present in the weak coupling
regime. This ghost is a radion, in a complete analogy with what happens in
the GRS model \cite{Pilo:2000et}. The difference is that the GRS radion is
 a ghost because the tension of the TeV brane is negative, while in our
case it is the presence of the induced term that makes the radion to be a
ghost.

The structure of this paper is as follows. In section \ref{main} we
describe the setup. In section \ref{linear} we present a simple technique,
based on the study of the junction conditions on the brane, which enables
one to single out the strong coupling domains in the parameter space. We
demonstrate that in the model under consideration there are two  strong
coupling domains separating three weak coupling regions. In one of the
latter regions the induced term is localized on the Planck brane and
gravity is exactly localized. In the other two regions the induced term is
on the TeV brane. Then one may expect modification of gravity at large
distances. However, in section \ref{radion} we demonstrate that the radion
is a ghost in these regions. We present our conclusions in section
\ref{conc}.

\section{The model}
\label{main}
\begin{figure}[t]
\begin{center}
\begin{picture}(500,150)
(20,0)
\put(20,0){ \epsfig{file=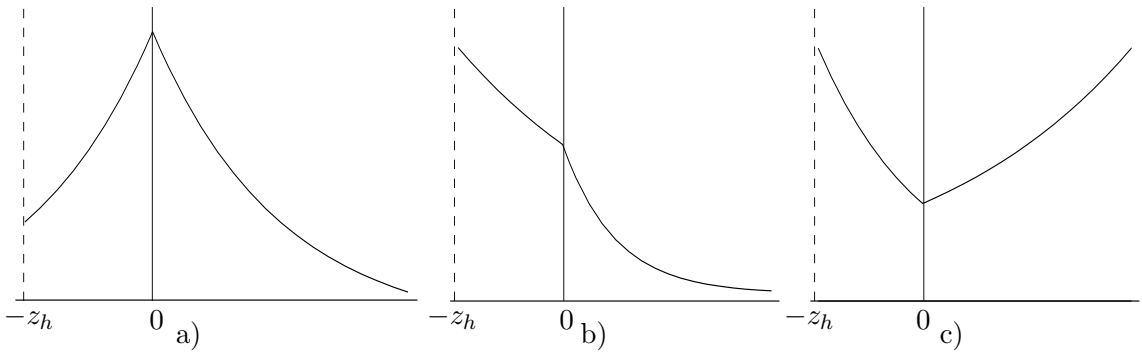,width=15cm,height=4cm} }
\put(20,-8){$-z_h$}
\put(75,-10){$0$}
\put(180,-8){$-z_h$}
\put(230,-10){$0$}
\put(315,-8){$-z_h$}
\put(365,-10){$0$}
\put(85,-15){a)}
\put(238,-15){b)}
\put(374,-15){c)}
\end{picture}
\end{center}
\caption{ Profiles of the warp factor $a(z)$ for different choices of
signs for $k_L$ and $k_R$: a) $k_L<0$, $k_R>0$ b) $k_L>0$, $k_R>0$ c)
$k_L>0$, $k_R<0$. Solid and dashed vertical lines show the position of the
visible brane and the hidden brane, respectively.}
\label{profiles}
\end{figure}

We consider the five-dimensional background with the metric
\be
\label{metric}
ds^2=a^2(z)\eta_{\mu\nu}dx^\mu dx^\nu-dz^2 \ee We assume
that this metric is a solution of the five-dimensional gravity with
three-brane sources, \bea S&=&-M^3\int d^5 X \sqrt{g} R^{(5)}-\Lambda\int
d^5x\sqrt{g}-\nonumber\\
&\sum_i& \lambda_i\int d^5 X \sqrt{-\gamma^{(i)}}\delta(z-z_i)
\label{action}
\eea where $g$ is the bulk metric and $\gamma^{(i)}$ is the
metric induced on the $i$-th brane. The five-dimensional cosmological
constant $\Lambda< 0$ may be different in different domains of the bulk
space, separated by three-branes. We consider the Lykken--Randall model
with two branes. The first (``hidden'') brane is placed at the fixed point
$z=-z_h$ of the $Z_2$ orbifold symmetry $z\to -2z_h-z$. The $z$ coordinate
of the second (``visible'') brane is $z=0$. Also we add to the action the
induced Einstein term \be
\label{indterm}
S_{ind}=-M^{2}_{ind} \int d^5 X
\sqrt{-\gamma^{(v)}}\delta(z)R^{(4)}(\gamma^{(v)}) \ee localized on the
visible brane. The explicit shape of the warp factor $a(z)$ is \be
a(z)=\left\{
\begin{array}{l}
\e^{-k_Lz}\;,\;\;-z_h<z<0\\
\e^{-k_Rz}\;,\;\;0<z
\end{array}
\right. \ee where the values of the inverse adS radius  on the left and on
the right of the visible brane $k_{L,R}$ are related to the values of the
cosmological constant on the left and on the right of the visible brane
$\Lambda_{L,R}$ as follows, \be
\label{ks}
|k_{L,R}|=\sqrt{-{\Lambda_{L,R}\over 12M^3}}\;. \ee The brane
tensions should satisfy the fine-tuning conditions \be
\label{finetune}
\lambda_h=12M^3k_L\;,\;\;\lambda_v=6(k_R-k_L)M^3\;. \ee
The profile of the warp factor $a(z)$ for different choices of sign in Eq.
(\ref{ks}) is shown in Fig. \ref{profiles}.
In what follows we do not consider the case $k_R<0$ in detail. In this
case one would have to introduce the third brane to screen the adS
boundary at $z=+\infty$; as we will comment later on, we do not expect any
qualitatively new features in that case.

Let us briefly recall the main properties of the two-brane model without
induced Einstein term for different values of the parameters $k_L$ and
$k_R$.

If $k_L<0$ and $k_R>0$ one can push the hidden brane to infinity. Then at
$k_L=-k_R$ one has just the second Randall--Sundrum model, while in
general  one arrives at the $Z_2$-asymmetric generalization of that model.
Due to the presence of a graviton zero mode localized on the visible
 brane, one recovers the conventional four dimensional gravity  at large
distances along this brane. The value of the four-dimensional Planck mass
is \be
\label{effplanck}
M_{0}^2=M^3\l {1\over k_R}-{1\over k_L}\r \ee At
distances shorter than $r_b=\max\{k_L^{-1},-k_R^{-1}\}$ bulk corrections
are essential, while quantum gravity effects are important at even shorter
distances of order $M^{-1}$.

If both $k_L$ and $k_R$ are positive then one cannot push the hidden brane
to infinity. This brane is needed to screen the adS boundary at
$z=-\infty$. In this case the wave function of the graviton zero mode is
peaked near the hidden brane, so the effective Planck mass on the visible
brane depends exponentially on the separation between the branes, \be
\label{effplanck1}
M_{0}^2=M^3\l {1\over k_R}+{1\over
k_L}\l\e^{2z_vk_L}-1\r\r \ee The long-distance potential $V(r)$ between
two masses $m_1$ and $m_2$ on the visible brane at the distance $r$ is
given by\footnote{Strictly speaking, this result was obtained in the case
of the zero tension brane $k_L=k_R$. Corrections to Eq. (\ref{mmLR})
present at $k_L\neq k_R\equiv k$ do not affect our discussion
here.}\cite{Lykken:1999nb,Giddings:2000mu} \be
\label{mmLR}
V(r)=\frac{1}{M_0^2}\frac{1}{r}\left(1+\frac{M^3}{M_0^2k^3r^2}
+\frac{M_0^2}{k^5M^3r^6} \right)\;.
\ee
Consequently, the four-dimensional
Newton's law does not hold at distances shorter than
\[
r_b=\frac{M_0^{1/3}}{k^{5/6}M^{1/2}}\;.
\]
Note, that this distance becomes larger when the distance between the
branes grows.

One more difference with the first case is that one does not recover the
standard Einstein gravity on the visible brane at long distances. Instead,
one has a scalar-tensor theory of gravity due to the presence of the
radion field.

Another feature of this model is that if $k_R<k_L $, then the tension of
the visible brane is negative (see Eq. (\ref{finetune})) and the radion
becomes a ghost \cite{Pilo:2000et}. For $k_R=0$ one recovers the GRS model
with quasilocalized gravity and ghost field.

The purpose of this paper is to study the modification of the above
results due to the presence of the induced term (\ref{indterm}) and in
particular to check that there is no corner of the parameter space with
long distance modification of gravity and without ghosts or strong
coupling at unacceptably low energy scale.

For the moment let us ignore the fact that the graviton is a tensor
particle and briefly discuss the effect one would expect from the induced
term in the scalar case (more detailed discussion of the scalar propagator
in the Randall--Sundrum model with induced term can be found in Ref.
\cite{Kiritsis:2002ca}). The brane-to-brane scalar propagator $G(p)$ in
the presence of the induced term is \be
\label{scalar}
G(p)={1\over M^3\l G_0^{-1}+r_cp^2\r} \ee where $G_0(p)$ is
the brane-to-brane propagator without induced term and
\[
r_c={M_{ind}^2\over M^3}
\]
At distances shorter than the adS radius one has $G_0(p)=1/p$, so that
at short distances the propagator $G(p)$ always has a four-dimensional
form with the Planck mass equal to $M_{ind}$. At very large distances one
again recovers four-dimensional propagator with Planck mass \be
\label{longM}
M_l^2=M_{ind}^2+M_0^2\; \ee So in principle one can have
modification of gravity at long distances in this model. Namely, at
relatively ``short'' distances
one expects 4d gravity with Planck mass $M_{ind}$, then a region of scales
where bulk contribution dominates, and 4d gravity with Planck mass $M_{l}$
at longer scales. Arranging the parameters in such a way that the first of
these transitions occurs at the cosmological scale, one would hope to
obtain interesting long-distance modification of gravity. Furthermore, for
positive $k_L$ and large enough distance $z_h$ between the branes, the
second of these transitions occurs at the length scale longer than the
curvature length in the bulk. Unfortunately, as we will see in the rest of
the paper, whenever large distance modification of gravity takes place,
there is either ghost or strong coupling at unacceptably low energy scale.
\section{Junction conditions and strong coupling}
\label{linear}
In this paper we study the effect of the induced term
(\ref{indterm}) in the regime when the parameter $M_{ind}$ is the largest
energy scale involved in the problem,
\[
M_{ind}\gg k_L\;,\;k_R\;,\;M\;.
\]
In flat space model with brane-induced gravity the following peculiar
effect happens \cite{Luty:2003vm,Rubakov:2003zb}. Some couplings
determining the interaction strength of the longitudinal (from 4d point of
view) components of the graviton involve positive powers of the parameter
$M_{ind}$. As a result, a new dynamical scale
\[
\Lambda= {M^2\over M_{ind}}
\]
emerges. Above this energy scale the theory is strongly coupled. In
what follows, we refer to this situation as strong coupling regime. It is
shown in Ref. \cite{Luty:2003vm}, that, unlike the case of flat bulk, in
the second Randall--Sundrum model with the induced term the strong
 coupling regime does not occur if
\[
M_{ind}^2>{M^3\over k}
\]

In this section we describe a simple technique to single out regions of
the parameter space where strong coupling regime takes place. A
straightforward way \cite{Rubakov:2003zb} to see the strong coupling  is
to calculate the full propagator in the theory. If this propagator
contains ``large'' terms proportional to positive powers of $M_{ind}$
which are not pure gauge everywhere, then strong coupling regime takes
place. A shortcut which we make use of here is the observation that these
terms show up already in the junction conditions for the metric
perturbations on the visible brane. The first junction condition we employ
is the Israel condition \cite{Israel:rt} for the discontinuity of the
extrinsic curvature $K_{\mu\nu}$. In the presence of the induced term this
condition takes the following form \be
\label{Israel}
\left[K_{\mu\nu}\right]^+_-={1\over M^3}\l -{1\over
6}\lambda\gamma^{(v)}_{\mu\nu}+{1\over2}\l\tau_{\mu\nu}-{1\over
3}\gamma^{(v)}_{\mu\nu}\tau\r-M_{ind}^2\l R_{\mu\nu}^{ind}-{1\over
6}\gamma^{(v)}_{\mu\nu}R^{ind}\r\r \ee where $\tau_{\mu\nu}$ is  the
energy-momentum tensor of matter residing on the brane ($\tau_{\mu\nu}$
does not include the brane tension).

To obtain the second equation, let us present the projection of the
Einstein tensor on direction normal to the brane in the following
 form\footnote{Here Latin indices $A,B$ take values $\mu, z$, while Greek
indices $\mu,\nu$ correspond to the coordinates tangent to the brane.} \be
\label{projrel}
G_{AB}n^An^B=R_{ABCD}\l g^{AC}+n^An^C\r\l g^{BD}+n^Bn^D\r
\ee where $R_{ABCD}$ is  the bulk Riemann tensor and $n^A$ is the unit
vector normal to the brane. Now, using the Gauss--Codacci relation, one
may express the quantity in the r.h.s. of Eq. (\ref{projrel}) through the
induced and extrinsic curvatures as follows \be
\label{GaussCod}
R_{ABCD}\l g^{AC}+n^An^C\r\l g^{BD}+n^Bn^D\r=R^{ind}+\l
K^\mu_\mu\r^2- K_{\mu\nu}K^{\mu\nu} \ee Finally, taking the component of
the Einstein equations normal to the brane and using Eqs. (\ref{projrel}),
(\ref{GaussCod}) , one gets \be
\label{constraint}
(K^\mu_\mu)^2-K_{\mu\nu}K^{\mu\nu}+{\Lambda\over
M^3}-{T_{AB}n^An^B\over M^3}+R^{ind}=0 \ee where the $T_{AB}$ is the
energy-momentum tensor of the bulk matter. It is worth pointing out, that
in the initial value formulation of general relativity as a constrained
Hamiltonian system, with the induced metric $\gamma_{\mu\nu}$ and
extrinsic curvature $K_{\mu\nu}$ being the dynamical coordinates and
momenta, Eq.~(\ref{constraint}) plays the role of the constraint on their
initial values (see, e.g., Ref. \cite{Wald:rg}). It is worth noting that
the constraint (\ref{constraint}) is valid on both sides of the brane
separately, so it actually provides {\it two} equations.

The strong coupling phenomenon happens in the scalar sector, so let us
consider the trace of the Israel condition (\ref{Israel}). Also, let us
decompose the extrinsic curvature as follows
\[
K_{\mu\nu}={a'\over a}\gamma_{\mu\nu}^{(v)}+\kappa_{\mu\nu}\;.
\]
For the background metric (\ref{metric}) one has $\kappa_{\mu\nu}=0$,
$\gamma_{\mu\nu}^{(v)}=\eta_{\mu\nu}$ and the junction conditions
(\ref{Israel}), (\ref{constraint}) imply the fine-tuning relations
(\ref{ks}), (\ref{finetune}). To the leading order in the deviations from
the background, one has from Eqs. (\ref{Israel}), (\ref{constraint})
\begin{eqnarray}
\left[\kappa\right]^+_- &=&-{1\over M^3}\l{1\over 6}\tau+{1\over
3}M_{ind}^2R^{(ind)}\r\nonumber\\
-6{a'\over a}\kappa&+&{T_{zz}\over M^3}=R^{(ind)}
\end{eqnarray}
where $\kappa\equiv \eta^{\mu\nu}\kappa_{\mu\nu}$. Again, we actually have
three equations, and using them one can find the values of $\kappa_{R(L)}$
on both sides of the brane and the induced scalar curvature $R^{(ind)}$.
The result is
\begin{eqnarray}
R^{(ind)}&=&{1\over
(k_R-k_L)M^3-2k_Rk_LM_{ind}^2}\l(k_R-k_L)T_{zz}+{1\over 2}\tau\r
\label{R}\\
\kappa_{R(L)}&=&{k_{L(R)}\over (k_R-k_L)M^3-2k_Rk_LM_{ind}^2} \l
{M_{ind}^2\over 3 M^3}T_{zz}+{1\over 6}\tau\r
\label{kappa}
\end{eqnarray}
Eq. (\ref{R}) implies that the induced curvature scalar never contains
large contributions proportional to the positive powers of $M_{ind}$. This
is the reflection of the fact that large terms in the full propagator are
pure gauge from the 4d point of view (cf. Refs.
\cite{Luty:2003vm,Rubakov:2003zb}). On the other hand, there is a
potentially large term proportional to $T_{zz}$ in the expression for the
trace of the extrinsic curvature. If \be
\label{strongcoupling}
r_c={M_{ind}^2\over M^3}\gg \left|{1\over
k_L}-{1\over k_R}\right| \ee the factor of $M_{ind}^2$ in this term
cancels out and there is no new strong coupling scale suppressed by
$M_{ind}^{-1}$. However, in the opposite case, the expression for the
extrinsic curvature $\kappa$ contains a term proportional to $M_{ind}^2$;
consequently, a term enhanced by $M_{ind}^2$ is  present in the Green's
 function and the strong coupling regime occurs.
\EPSFIGURE[ht]{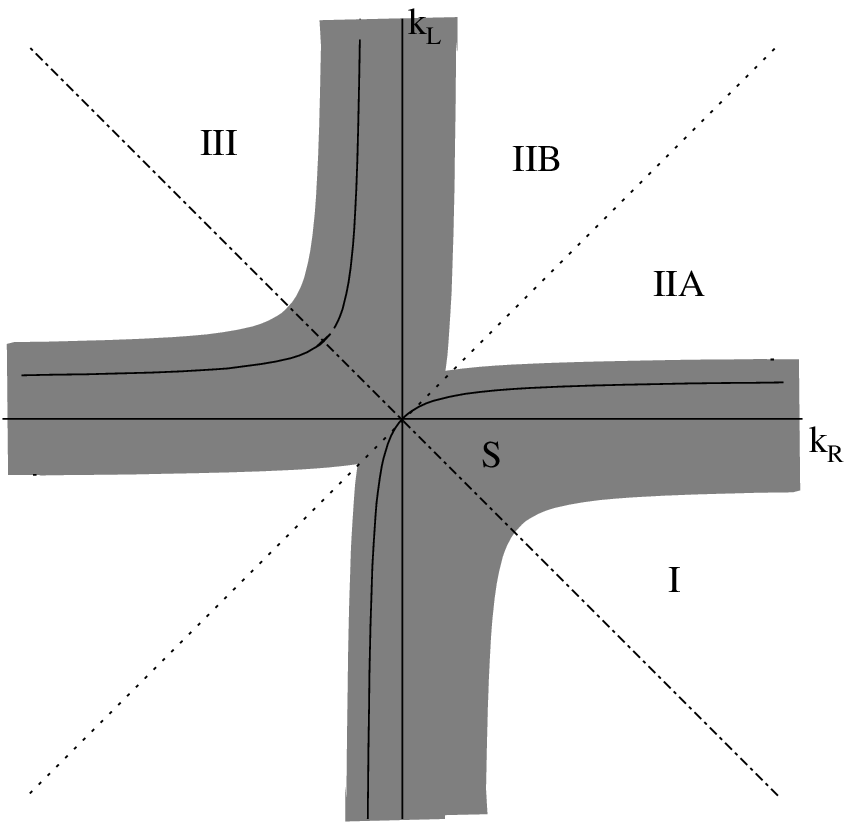}{Parameter space of the
two brane model with induced term. Strong coupling occurs in the grey
shaded region $S$. Solid lines in this region are lines of zeroes of the
denominator in Eq. (\ref{kappa}). Below the dotted line both branes have
positive tensions. On the dash-dotted line one can perform $Z_2$
identification on the visible brane and arrive at the Randall--Sundrum
model.
\label{diagramm}
}
In Fig. \ref{diagramm} the plane of the parameters $k_L$ and $k_R$ is
shown. In the shaded grey region one has
\[
r_c\lesssim \left|{1\over k_L}-{1\over k_R}\right|
\]
so that strong coupling phenomenon happens there. The physical meaning
of this region is clear: here the distance $r_c$ is smaller then the adS
radius on one of the sides of the visible brane. Consequently, the strong
coupling is nothing but the same phenomenon as observed in the flat model.
One can estimate the scale $\Lambda$ of the strong coupling in the
following way. Let us, for definiteness, consider the case $k_R>k_L$. From
Eq. (\ref{kappa}) one finds that on the left of the visible brane one has
\[
\kappa_L\sim {M_{ind}^2\over M^6} T_{zz}
\]
Consequently, the graviton propagator contains a large term
proportional to $M_{ind}^2/M^6$. Now, the three-graviton vertex becomes
large if
\[
M^3\l { M_{ind}^2\over M^6}\r^{3/2}E^3>1\;.
\]
where a factor of $M^3$ comes from the vertex\footnote{Note that, as we
discussed above, the large term in the graviton propagator is pure gauge
from the 4d point of view, so it cancels in the vertices coming from the
brane-induced action.} and the power of energy $E$ is restored on
dimensional grounds. This gives
\[
\Lambda\sim {M^2\over M_{ind}}
\]
in agreement with the analysis in the flat case.

In the region $I$ in Fig. \ref{diagramm} one has $k_L<0$, the induced term
is on the Planck brane so one can remove the hidden brane. From the
comparison of Eqs. (\ref{strongcoupling}), (\ref{longM}) and
(\ref{effplanck}) it follows that the absence of the strong coupling
implies here that the Planck masses in the UV and IR are both
approximately equal to $M_{ind}$, so no large scale  modification of
gravity happens. Note, however, that the quantum gravity scale $\Lambda$
is changed as compared to the case $M_{ind}=0$ (when it is equal to $M$)
even in this region. Let us consider, for example, the $Z_2$ symmetric
case $|k_L|=k_R=k$. Then it follows from Eq. (\ref{kappa}) that the
graviton propagator contains a term proportional to $1/(M^3k)$ and the
same analysis as above implies that
\[
\Lambda\sim\sqrt{Mk}
\]
in agreement with the result in Ref. \cite{Luty:2003vm} for the
Randall--Sundrum model with a single brane.

The analysis of the region $I$ remains valid if the hidden brane is at
finite distance $z_h$. Again, the 4d Planck mass is approximately equal to
$M_{ind}$, there is no long distance modification of gravity, but the
strong coupling scale is below the 4d Planck mass and fundamental mass
$M$.

As we discussed above, in the region $IIA$ in Fig. \ref{diagramm} both
branes have positive tensions and, at large enough separation between the
branes,  one may expect long distance modification of gravity. However, as
we show below, radion is a ghost in the whole region $II$.
\section{Effective action for the radion}
\label{radion}
In this section we consider the region $k_L>0$. We
calculate the quadratic effective action for the radion field and
demonstrate that the radion is a ghost if \be
\label{ghostregion}
M_{ind}^2>{ M^3\over 2}\l{1\over k_L}-{1\over k_R}\r
\ee i.e., above the right solid line in Fig. \ref{diagramm}. The
calculation of the quadratic effective action is very similar to that in
the case of the Lykken--Randall model without induced term
\cite{Pilo:2000et}. First, following Ref. \cite{Charmousis:1999rg}, we
calculate the wave function of the radion in two different patches,
corresponding to the Gaussian normal coordinates with respect to the
hidden and visible branes, so that the metric has the form
\[
ds^2=a^2(z)(\eta_{\mu\nu}+h^{(i)}_{\mu\nu})dx^\mu dx^\nu-dz^2
\]
where $i=\alpha,\beta$ labels different patches. Then we perform a
gauge transformation to a single patch and plug the resulting wave
function into the quadratic action.

The calculation of the radion wave function is completely parallel to that
in Ref. \cite{Charmousis:1999rg}, so we just present the results. In the
first patch, that covers the region $-z_h\leq z< 0$, one has \be
\label{firstpatch}
h^{(\alpha)}_{\mu\nu}=f\eta_{\mu\nu}-\d_\mu\d_\nu
f{\e^{2k_Lz}\over 2k_L^2}+ \d_\mu\d_\nu f {\e^{2k_L(2z+z_h)}\over 4k_L^2}
\ee Here $f(x)$ is the (unnormalized) radion field. The on-shell condition
for the radion is $\d_\mu\d^\mu f=0$. In the second patch, that covers the
region $-z_h<z<\infty$ one has
\begin{eqnarray}
h^{(\beta)}_{\mu\nu}= \left\{
\begin{array}{l}
C \l \displaystyle f\eta_{\mu\nu}-\d_\mu\d_\nu f{\displaystyle
\e^{2k_Lz)}\over 2k_L^2}\r+\displaystyle \d_\mu\d_\nu f {\displaystyle
\e^{2k_L(2z+z_h)}\over 4k_L^2}\;\; {\rm for }\;\; z<0\\
C \l \displaystyle f\eta_{\mu\nu}-\d_\mu\d_\nu f{\displaystyle
\e^{2k_Rz}\over 2k_R^2}\r+\l \displaystyle{C\over 2}\l{1\over
k_R^2}-{1\over k_L^2}\r+{\displaystyle \e^{2k_Lz_h}\over 4k_L^2}\r
\d_\mu\d_\nu f \;\; {\rm for }\;\; z>0
\label{Arr/Pg12/1:paper}
\end{array}
\right.
\end{eqnarray}
where
\[
C={\e^{2k_Lz_h}k_R\over k_R-k_L-2M_{ind}^2k_Rk_L/M^3}\;,
\]
$C<0$ in the regions $IIA$ and $IIB$ in Fig.~\ref{diagramm}.
Note, that the wave function (\ref{Arr/Pg12/1:paper}) is pure gauge for
$z>0$, so the growth of $h^{(\beta)}_{\mu\nu}$ toward $z=+\infty$ is not
dangerous. To calculate the effective action for the radion field let us
perform gauge transformation to the coordinate system covering the whole
space. This gauge transformation can be performed in two steps. First, one
makes the shift of $z$-coordinate in the second patch \be
\label{firsthift}
z\to z-{\l C-1\r\over 2k_L}f\;. \ee The resulting metric
is described by a single patch, however, the visible brane is bended in
this coordinate system. To eliminate this bending one performs the second
gauge transformation
\begin{eqnarray}
&z\to& z+ { C-1\over 2k_L}f B(z)\nonumber\\
&x^\mu\to& x^\mu+\xi^\mu(x,z)
\label{secondshift}
\end{eqnarray}
where $B(z)$ is an arbitrary function satisfying
\[
B(0)=0\;,\;\; B(z_v)=1
\]
and functions $\xi^\mu(x,z)$ are fixed by the requirement that the
$(\mu z)$-components of the resulting metric vanish. Explicitly one has
\[
\xi^\mu=-a^2{C-1\over 2k_L}B(z)\eta^{\mu\nu}\d_\nu f\;.
\]
The resulting metric has the form
\[
ds^2=a^2(z)(\eta_{\mu\nu}+h_{\mu\nu})d x^\mu d x^\nu-(1-h_{zz})dz^2
\]
where
\begin{eqnarray}
h_{zz}&=&{C-1\over 2k_L}fB'(z)
\label{rzz}\\
h_{\mu\nu}&=&h_{\mu\nu}^{(\alpha)}-{\e^{-2k_L(z-z_v)}(C-1)\over
2k_L}\d_\mu\d_\nu f\int dz B(z)\e^{2k_L(z-z_v)}+2(C-1)\eta_{\mu\nu}fB(z)
\label{rmn}
\end{eqnarray}
for $-z_h\leq z\leq 0$ and pure gauge for $z>0$. To calculate the
effective action for the radion we plug expressions (\ref{rzz}) and
(\ref{rmn}) into the quadratic action presented in the form
\[
S_2=-{1\over 2}\int dzd^4x\sqrt{g}\delta g^{AB}E_{AB}[\delta g]
\]
and take the integral over $z$-coordinate. Here tensor $E_{AB}[\delta
g]$ is the linear part of the variation of the action given by Eqs.
(\ref{action}) and (\ref{indterm}). As in the case of the Lykken--Randall
model without induced term \cite{Pilo:2000et}, only $(zz)$-components of
the tensor $E_{AB}[\delta g]$ are non-zero. Finally, one arrives at the
following effective action for the radion
\begin{gather}
\label{result}
S_{rad}=-3\e^{-2k_L z_v }M^3{C-1\over
2k_L}\int_0^{z_v}dzB'(z)\int d^4xf\d_\mu^2 f\\
=-{3\over 2}\e^{-2k_L z_v }{M^3\over k_L}\l {\e^{2k_Lz_v}k_R\over
k_R-k_L-2M_{ind}^2k_Rk_L/M^3}-1\r\int d^4xf\d_\mu^2 f
\end{gather}
We see, that the radion is indeed a ghost in the regions $IIA$ and $IIB$
in Fig. \ref{diagramm}. \section{Concluding remarks}
\label{conc}
To conclude, let us first comment on what happens in the
region $III$ which was not discussed so far. When $k_R$ becomes negative,
 the warp factor $a(z)$ has the shape shown in Fig.~\ref{profiles}c) and
the third brane is needed to screen the adS boundary at $z=+\infty$.
Consequently, the second radion appears here. This radion becomes a ghost
above the left solid line in Fig. \ref{diagramm}. Consequently, there are
two ghost fields in the region $III$. On the dash-dotted line one can make
$Z_2$ identification with respect to the visible brane. This
identification removes one of the ghosts in the region $III$, but the
second one is still there in agreement with Ref. \cite{Luty:2003vm}.

To summarize, the results of this paper suggest that it is unlikely to
cure the problems of the brane-induced gravity models by invoking branes
of non-zero tension and warped bulk space. These results as well as those
obtained in Refs.
\cite{Pilo:2000et,Dubovsky:2002jm,Luty:2003vm,Rubakov:2003zb} indicate
that non-locality in the low-energy effective theory arising due to the
presence of infinitely large extra dimensions is not sufficient to get
around the no-go results of Ref. \cite{Arkani-Hamed:2002sp} and some more
radical approach is needed to achieve large scale modification of gravity
in the consistent theory.

\section*{Acknowlegements} We would like to
thank Yu.~Kubyshin, D.~Levkov,  M.~Shaposhnikov,
S.~Sibiryakov, P.~Ti\-nya\-kov, and especially V.~Rubakov and R.~Rattazzi
for helpful discussions. S.D. would like to thank the IPT at the Lausanne
Univ., where part of this work was done, for kind hospitality.

This work is supported in part by the IISN (Belgium), the ``Communaut\'e
Fran\c{c}aise de Belgique''(ARC), and the Belgium Federal Government
(IUAP) (M.L.); by RFFI grant 02-02-17398; by the Grants of the President
of the Russian Federation NS-2184.2003.2; by the programme SCOPES of the
Swiss NSF, project No.~7SUPJ062239, financed by Federal Department of
Foreign affairs.

\end{document}